\documentclass[fleqn,11pt,twoside]{article}

\usepackage{amsthm,amsthm,amssymb, color, xcolor,epsfig, graphics, subfigure}

\usepackage{amsmath, graphicx, latexsym, lscape}

\makeatletter
\newcommand{\copyrightnote}[2]{{\renewcommand{\thefootnote}{}
 \footnotetext{\small\it
\begin{flushleft}
 \copyright \ #1   #2  
\end{flushleft}}}}

\newcommand{\Name}[1]{\begin{flushleft}
                       \LARGE \bf #1
                       \end{flushleft}\vspace{-3mm}}

\newcommand{\Author}[1]{\begin{flushleft}
                       \it #1 \end{flushleft}}

\newcommand{\Address}[1]{\begin{flushleft}
                       \it #1 \end{flushleft}}

\newcommand{\Date}[1]{\begin{flushleft}
                      \small  \it #1 \end{flushleft}}

\newcommand{\evenhead}{Author \ name}
\newcommand{\oddhead}{Article \ name}

\renewcommand{\@evenhead}{
\hspace*{-3pt}\raisebox{-15pt}[\headheight][0pt]{\vbox{\hbox to \textwidth
{\thepage \hfil \evenhead}\vskip4pt \hrule}}}
\renewcommand{\@oddhead}{
\hspace*{-3pt}\raisebox{-15pt}[\headheight][0pt]{\vbox{\hbox to \textwidth
{\oddhead \hfil \thepage}\vskip4pt\hrule}}}
\renewcommand{\@evenfoot}{}
\renewcommand{\@oddfoot}{}

\long\def\@makecaption#1#2{%
  \vskip\abovecaptionskip
  \sbox\@tempboxa{\small \textbf{#1.}\ \ #2}%
  \ifdim \wd\@tempboxa >\hsize
    {\small \textbf{#1.}\ \ #2}\par
  \else
    \global \@minipagefalse
    \hb@xt@\hsize{\hfil\box\@tempboxa\hfil}%
  \fi
  \vskip\belowcaptionskip}

\newcommand{\JNMPnumberwithin}[3][\arabic]{%
  \@ifundefined{c@#2}{\@nocounterr{#2}}{%
    \@ifundefined{c@#3}{\@nocnterr{#3}}{%
      \@addtoreset{#2}{#3}%
      \@xp\xdef\csname the#2\endcsname{%
        \@xp\@nx\csname the#3\endcsname .\@nx#1{#2}}}}%
}

\newcommand{\resetfootnoterule} {
  \renewcommand\footnoterule{%
  \kern-3\p@
  \hrule\@width.4\columnwidth
  \kern2.6\p@}
}

\renewcommand{\footnoterule}{}

\makeatother

\theoremstyle{definition}

\def \D {\hbox{d}}
\def \tr{\mathop{\rm tr}\nolimits}
\newcommand\fg{\mathfrak{g}}

\begin{document}

\renewcommand{\evenhead}{ {\LARGE\textcolor{blue!10!black!40!green}{{\sf \ \ \ ]ocnmp[}}}\strut\hfill 
Robert Conte
}
\renewcommand{\oddhead}{ {\LARGE\textcolor{blue!10!black!40!green}{{\sf ]ocnmp[}}}\ \ \ \ \  
On a reduction of the Drinfeld-Sokolov hierarchy
}

\thispagestyle{empty}
\newcommand{\FistPageHead}[3]{
\begin{flushleft}
\raisebox{8mm}[0pt][0pt]
{\footnotesize \sf
\parbox{150mm}{{Open Communications in Nonlinear Mathematical Physics}\ \ \ \ {\LARGE\textcolor{blue!10!black!40!green}{]ocnmp[}}
\quad Special Issue 2, 2024\ \  pp
#2\hfill {\sc #3}}}\vspace{-13mm}
\end{flushleft}}

\FistPageHead{1}{\pageref{firstpage}--\pageref{lastpage}}{ \ \ }

\strut\hfill

\strut\hfill

\copyrightnote{The author(s). Distributed under a Creative Commons Attribution 4.0 International License}

\begin{center}

{\bf {\large Proceedings of the OCNMP-2024 Conference:\\ 

\smallskip

Bad Ems, 23-29 June 2024}}
\end{center}

\smallskip

\Name{On an equation arising by reduction of the Drinfeld-Sokolov hierarchy}

\Author{Robert Conte$^{\,1,2}$}

\Address{$^{1}$ Centre Borelli, \'Ecole normale sup\'erieure de Paris-Saclay, France.\\[2mm]
$^{2}$ Department of Mathematics, The University of Hong Kong, Hong Kong}

\Date{Received April 18, 2024; Accepted May 8, 2024}

\setcounter{equation}{0}

\begin{abstract}

\noindent 
A seventh order ordinary differential equation (ODE) arising by reduction of the Drinfeld-Sokolov hierarchy
is shown to be identical to a simi\-larity reduction of an equation
in the hierarchy of Sawada-Kotera.
We also exhibit its link with a particular F-VI,
a fourth order ODE isolated by Cosgrove which is likely to define 
a higher order Painlev\'e function.
\end{abstract}

\label{firstpage}


\section{Introduction}

In a recent article \cite{Liu-Wu-Zhang-JLMS},
the authors consider the tau cover of the Drinfeld-Sokolov hierarchy and,
in order to obtain explicit solutions,
perform a similarity reduction \cite[Eq.~(5.1)]{Liu-Wu-Zhang-JLMS}
which defines a system of nonlinear ODEs in the independent variable $x$.
By construction, this system 
possesses a Lax pair $(L,M)$ \cite[Eq.~(5.5)]{Liu-Wu-Zhang-JLMS}
whose zero-curvature condition is 
\begin{eqnarray}
& & 
[z \partial_z -M, \partial_x - L] \equiv z L_z - M_x +[L,M]=0,
\end{eqnarray}  
in which $z$ is the spectral parameter.

For all their choices but one of the underlying affine Kac-Moody algebra $\fg$,
the authors did succeed to explicitly integrate the nonlinear ODE system
in terms of various elliptic or Painlev\'e or higher Painlev\'e functions.
The only system which could not be integrated results from the choice $\fg=A_2^{(2)}$,
this is 
the seventh order nonautonomous system for $u(x),\omega(x)$ \cite[Example 5.5 page 1487]{Liu-Wu-Zhang-JLMS}
\begin{eqnarray}
& & {\hskip -11.0 truemm}
\left\lbrace
\begin{array}{ll}
\displaystyle{
\omega'-\frac{u}{3}=0,
}\\ \displaystyle{
u^{(6)} + 14 u u^{(4)} + 14 u' u^{(3)} + 14 {u''}^2 + 56 u^2 u'' + 28 u {u'}^2 + \frac{56}{3} u^4 + 36 x u +108 \omega=0.
}
\end{array}
\right.
\label{eqsys7}
\end{eqnarray}
which can be viewed as a birational transformation between $u(x)$ and $\omega(x)$,
each variable obeying a seventh order ODE.

The purpose of this work is to explicitly integrate this system,
i.e.~to map it 
either to Painlev\'e equations (second order),
or     to one of the five ``higher Painlev\'e equations'' (fourth and fifth order) isolated by Cosgrove \cite{CosPole2,CosPole1},
or     to higher order (six and above) equations in the hierarchy of the previous ones.

The method, developed in next sections, is classical and it
relies on three pieces of information: 
(i) the Lax pair, 
(ii) the singularity structure,
(iii) exhaustive lists of ODEs possessing the Painlev\'e property.

In Section \ref{sectionFirstIntegral},
by considering the invariants of the matrix Lax pair,
we obtain a unique first integral,
thus lowering the differential order only to six.
This is an indication (not a proof) that the equations of Painlev\'e and Cosgrove
should be insufficient to perform the integration.

In Section \ref{sectionSingularity},
we therefore investigate the singularity structure of the system (\ref{eqsys7}).
The three families of movable simple poles are then compared 
with sixth or senventh order members of various, already classified, hierarchies.
This allows one to integrate (\ref{eqsys7}) in terms of a higher member of the Sawada-Kotera hierarchy.

\section{First integral} 
\label{sectionFirstIntegral}

The system (\ref{eqsys7}) admits a three-dimensional zero curvature representation,
this is [courtesy of Wu Chao-Zhong],
\begin{eqnarray}
& & 
L=\sqrt{2} (-L_0+u L_1 -z L_2),
M=2 \sum_{j=0}^5 z^j M_j,
z L_z - M_x +[L,M]=0,
\nonumber\\ & & 
M_5=2 \sqrt{2} L_2,
M_4=2 \sqrt{2} L_0 - \frac{4}{3}\sqrt{2} u L_1,
M_3=\frac{2}{3} \sqrt{2} u M_{3a} +\frac{2}{3} u' M_{3b},
\nonumber\\ & & 
M_2=\frac{\sqrt{2}}{9} \left(u''+ 2 u^2\right) m_2,
\nonumber\\ & & 
M_1=\frac{u^{(3)} + 4 u u'}{9} M_{1a} + \frac{\sqrt{2}}{81} (3 u^{(4)} +9 {u'}^2- 8 u^3  - 27 x) L_2
\nonumber\\ & & 
M_0=A_1 M_{0a} +A_2 L_1 + A_3 L_0,
\nonumber\\ & & 
A_1=-\frac{u^{(5)} + 12 u' u'' +6 u u^{(3)} +8 u^2 +9 }{54},
\nonumber\\ & & 
A_2=  \frac{\sqrt{2}}{486} (20 u^4 + 60 u^2 u'' + 84 u {u'}^2 + 24 u u^{(4)} + 33 {u''}^2 + 60 u' u^{(3)} + 3 u^{(6)} + 108 x u  - 162 \omega ),
\nonumber\\ & & 
A_3=- \frac{\sqrt{2}}{81} (3 u^{(4)} + 18 u u'' +9 {u'}^2 - 8 u^3).
\label{eqZCC}
\end{eqnarray}  
in which the eight constant operators can be represented by third order matrices, 
\begin{eqnarray}
& & 
L_0   =\begin{pmatrix} 0 & 0 & 1 \cr   0 & 0 & 0  \cr   0 & 0 & 0  \cr \end{pmatrix},
L_1   =\begin{pmatrix} 0 & 0 & 0 \cr   0 & 0 & 0  \cr   1 & 0 & 0  \cr \end{pmatrix},
L_2   =\begin{pmatrix} 0 & 0 & 0 \cr   1 & 0 & 0  \cr   0 & 1 & 0  \cr \end{pmatrix},
\nonumber\\ & & 
M_{3a}=\begin{pmatrix} 0 & 1 & 0 \cr   0 & 0 & 1  \cr   0 & 0 & 0  \cr \end{pmatrix},
M_{3b}=\begin{pmatrix} 0 & 0 & 0 \cr   1 & 0 & 0  \cr   0 &-1 & 0  \cr \end{pmatrix},
m_{2} =\begin{pmatrix}-1 & 0 & 0 \cr   0 & 2 & 0  \cr   0 & 0 &-1  \cr \end{pmatrix},
\nonumber\\ & & 
M_{1a}=\begin{pmatrix} 0 &-1 & 0 \cr   0 & 0 & 1  \cr   0 & 0 & 0  \cr \end{pmatrix},
M_{0a}=\begin{pmatrix} 1 & 0 & 0 \cr   0 & 0 & 0  \cr   0 & 0 &-1  \cr \end{pmatrix}.
\end{eqnarray}  

The trace of $M^2$ is an affine function of $z^4$,
and the coefficient of $z^0$ is a single first integral $K$ of the system (\ref{eqsys7}),
\begin{eqnarray}
& & 
\tr M^2= K -2^4. 3^6 \omega z^4,\
\nonumber\\ & &
K= {v'}^2 + 4 v \left[2 u u^{(4)} -2 u' u^{(3)} + {u''}^2 + 12 u^2 u'' + 4 u^4 + 54 \omega\right],
\label{eqK}
\\ & & 
v= u^{(4)} + 6 u u''  + 3 {u'}^2 + \frac{8}{3} u^3 + 9 x.         
\nonumber
\end{eqnarray}
As to the traces of higher powers of $M$, all nonzero, they do not generate other first integrals.

The initial system (\ref{eqsys7}) is then equivalent to a birational transformation
between $u(x)$ and $v(x)$, each variable obeying a sixth order ODE,
\begin{eqnarray}
& & 
u^{(4)} + 6 u u'' + 3 {u'}^2 + \frac{8}{3} u^3 + 9 x -v=0,
2 v v'' - {v'}^2 + 8 u v^2 +K=0.
\label{eqsys6uv}
\end{eqnarray}
	
\section{Singularity structure} 
\label{sectionSingularity}

For the technical vocabulory, we refer for instance to Ref.~\cite{CMBook2}.
Near a movable singularity $x_0$, 
the system (\ref{eqsys7}) has three families of movable poles,
the first one being already mentioned in \cite[page 1487]{Liu-Wu-Zhang-JLMS}, 
and the seven Fuchs indices of the linearized equation near this singularity are all relative integers,
\begin{eqnarray}
& & 
\left\lbrace
\begin{array}{ll}
\displaystyle{
\omega \sim a \chi^{-1}, u \sim -3 a  \chi^{-2}, \chi=x-x_0,
}\\ \displaystyle{a=1, \hbox{ Fuchs indices= }       -1, 2, 3, 4, 7, 8,    12,
}\\ \displaystyle{a=2, \hbox{ Fuchs indices= } -2,   -1,    3, 4,    8, 9,     14,
}\\ \displaystyle{a=5, \hbox{ Fuchs indices= } -5,-4,-1,          7, 8,    12,     18.
}
\end{array}
\right.
\label{eqFamilies}
\end{eqnarray}

This structure matches that of a seventh order equation in the Sawada-Kotera hierarchy
mentioned by Gordoa and Pickering \cite[Eq.~(5.94)]{GP1999JMP}
(see also \cite{GP1999EPL}),
\begin{eqnarray}
& & {\hskip -11.0 truemm}
\left\lbrace
\begin{array}{ll}
\displaystyle{
\frac{\D}{\D X} \left[U^{(6)} + (7/2) (U U^{(4)} + U' U^{(3)} + {U''}^2 + U^2 U'') + 7/4 U {U'}^2 + 7/24 U^4\right]
}\\ \displaystyle{\phantom{1234}
-q_1 (2 U + X U')=0, q_1 = \hbox{ arbitrary nonzero constant},
}
\end{array}
\right.
\label{eqJMP594}
\end{eqnarray}
with the correspondence
\begin{eqnarray}
& & 
u=\frac{b^2}{4} U, X=b x, b^7=-\frac{36}{q_1}.
\end{eqnarray}
As to the scalar Lax pair \cite[Eqs.~(5.6), (5.96), (5.84)]{GP1999JMP},
as the interested reader can check,
it is identical to that obeyed by the second component of the wave vector of the matrices in (\ref{eqZCC}).


		
\section{A link to F-VI} 

As a final remark, let us mention a link between this seventh order equation in the Sawada-Kotera hierarchy
and the F-VI ODE isolated by Cosgrove \cite{CosPole2},
\begin{eqnarray}
& \hbox{(F-VI)} &
U''''=18 U U'' +9 {U'}^2 - 24 U^3 + \alpha U^2 + \frac{\alpha^2}{9} U
 + \kappa X + \beta.
\label{eqCosgroveFVI}
\end{eqnarray}

Indeed, when $K=0,v=0$, 
the system (\ref{eqsys6uv}) admits a particular solution $u(x)$ equal to 
a particular affine transform of F-VI
with the correspondence
\begin{eqnarray}
& & u= - 3 b^2 U, X=b\left(x-\frac{\beta}{3} b^6\right), \kappa=3 b^{-7}, \alpha=0.
\end{eqnarray}

\subsection*{Acknowledgements}

This is a pleasure to thank Zhang Youjin and Andrew Pickering for fruitful discussions,
and Wu Chao-Zhong for providing us with the precise matrix Lax pair.

\label{lastpage}
\end{document}